\documentclass[journal]{IEEEtran}
\usepackage{graphicx,subcaption}
\usepackage{cite}
\usepackage{url}     % adding urls like this: \url{http://www.stack.nl/~jwk/latex/}
\usepackage{soul}
\usepackage{verbatim}
\usepackage{times}		% Times font for the paper instead of default LaTeX fonts
\usepackage{multirow}	% Allow multiple-row elements in tabular environment
\usepackage [justification=centering]{caption}
\usepackage{amsmath}
\usepackage{amsfonts}
\usepackage{lipsum}
\usepackage{mathtools}
\usepackage{cuted}
\usepackage{hyphenat}	% Turn ON hyphenation to make text
\usepackage{lineno} % need \linenumbers command in text
\usepackage[usenames]{color} % provides colored text, e.g. for editorial memos
\usepackage{textcomp} % add. symbols like \textdegree \textohm \textCelsius
\usepackage[dvipsnames]{xcolor}
\definecolor{myblue}{rgb}{0.180, 0.453, 0.683}
\usepackage{comment}
\usepackage{xspace}

\newcommand{\ffr}{\ensuremath{{f_\mathit{fr}}}\xspace}

\newcommand{\finj}{\ensuremath{{f_\mathit{inj}}}\xspace}

\newcommand{\degree}{\ensuremath{{^\circ}}\xspace}

\begin{document}
\raggedbottom
% ========== %
%\linenumbers % uncomment to put linenumbers in the margin; this conflicts with the

% ========== %
\title{Phase Error Sensitivity to Injection Signals in Multi-Phase Injection-Locked  Ring Oscillators}
% ========== %
%
% ========== %
\author{	Zhaowen~Wang\vspace{-2em}
}%

\maketitle
% ========== %
%
% ========== %
\begin{abstract}
  Multi-phase injection-locked ring oscillators~(MP-ILROs) are widely used for multi-phase clock generation, with their phase accuracy primarily determined by the inherent accuracy of the oscillator itself, due to the suppression of input signal errors. However, a quantitative analysis of the oscillator's sensitivity to input errors remains largely unexplored. This paper presents a phasor-based analysis of injection locking, revealing that the phase error sensitivity is influenced by factors such as injection strength and the free-running frequency of the oscillator. Simulation results align closely with theoretical calculations, validating the effectiveness of the proposed method.  
\end{abstract}
%\vspace{-10pt}
% ========== %
%
% ========== %
%\vspace{-10pt}
% ========== %
%
%======================================================================
\section{Introduction}
\label{sec:intro}
%======================================================================

Driven by the ever-growing demand for bandwidth in large-scale computing and training platforms, wireline and optical links are rapidly increasing their data rates~\cite{Wang2022_thesis,Poon2021,Chen2018,Xie2024}. Multi-phase clock generation is a critical technique for enhancing transmitter serialization and receiver deserialization rates~\cite{Wang2021,Xu2024,Li2024}. Additionally, accurate phase shifts, achieved through phase interpolators utilizing multi-phase clock inputs, are essential for effective clock-data recovery~\cite{Chen2018,Wang2022,Mishra2022,Wang2022JSSC,Mohapatra2023,Mohapatra2024,Wang2024}. As symbol rates decrease, the importance of high-quality multi-phase clock generation in wireline clocking becomes paramount.

Multi-phase injection-locked ring oscillators (MP-ILROs) have gained prominence as an effective solution for multi-phase clock generation, thanks to their inherent symmetry and low jitter performance~\cite{Kinget2002,Wang2021JSSC,Xu2024,Li2024}. The multi-phase input effectively suppresses the noise of the ring oscillator, while the oscillator itself refines coarse input phases to generate precise multi-phase outputs. However, the mechanisms through which the ring oscillator mitigates input phase errors and its sensitivity to these errors remain insufficiently understood. While a high injection ratio helps reduce jitter in the ring oscillator, it can also compromise the phase improvement of the input signals. Therefore, developing an analytical model to quantify the phase improvement ratio is crucial for guiding design decisions and achieving a balance between jitter reduction and phase accuracy. Unfortunately, existing models including the time-domain analysis, phase-domain analysis and frequency-domain analysis, predominantly focus on locking range analysis or phase noise analysis, leaving phase accuracy analysis largely unexplored~\cite{Adler1946,Razavi2004,Gangasani2006,Yuan2012,Hong2019,Zhang2022,Wang2025}.

This work proposes an analytical framework for assessing phase error sensitivity using the phasor diagram of injection locking~\cite{Adler1946,Wang2025}. Input errors are modeled as small disturbances to the original phasor. By analyzing the relationship between these disturbances and the phasor, the phase error sensitivity can be derived. Simulation results are compared with theoretical calculations from the proposed model, demonstrating strong agreement and validating the model’s accuracy.

This paper is organized as follows: Section~\ref{sec:RO Model} describes the model that analyzes the MP-ILRO phase accuracy compares the analytical and simulation results. Section~\ref{sec:conclusions}  concludes the paper.

%======================================================================
\section{CMOS Ring Oscillator Modeling}
\label{sec:RO Model}
%======================================================================

\begin{figure}[t!]
\centering 
\includegraphics[width=0.48\textwidth]{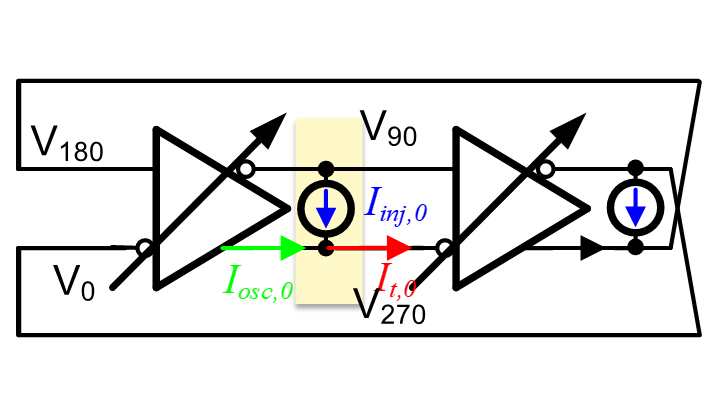}
\vspace{-6pt}
\caption{\small{The schematic diagram of an injection-locked ring oscillator with multi-phase injection.}}
\label{fig:ro}
\vspace{-12pt}
\end{figure}

Fig.\ref{fig:ro} illustrates the basic operation of a two-stage differential injection-locked oscillator~(ILO). The injection signal can be modeled as a current source, $I_{inj}$ , flowing into one of the oscillation nodes. In a multi-phase injection-locked configuration, phase-shifted injection signals flow into each of the corresponding injection nodes, e.g. $I_{inj,0}$ and $I_{inj,90}$.

%----FIGURE-----
%---------------

Due to the high-frequency operation of the ring oscillator, its voltage and current waveforms are primarily dominated by the fundamental tone~\cite{Zhang2022,Wang2025}. Therefore, a phasor model remains valid for analyzing injection locking, encompassing parameters such as locking range, phase accuracy, and phase noise. For CMOS-based oscillators, the amplitude and phase shift are highly sensitive to the input signal's amplitude, necessitating consideration of amplitude-to-phase conversion effects within the delay stage~\cite{Kabbani2003,Wang2021JSSC,Wang2025}.

\begin{figure}[t!]
\centering 
\includegraphics[width=0.35\textwidth]{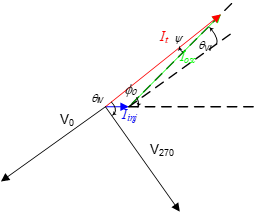}
\vspace{-6pt}
\caption{\small{The phasor diagram of the voltages and currents in an injection-locked ring oscillator.}}
\label{fig:phasor}
\vspace{-12pt}
\end{figure}

The phasor construction is depicted in Fig.~\ref{fig:phasor}. Here, $V_0$ represents the 0~\degree voltage waveform, which triggers the oscillator current $I_{osc,0}$ with a phase shift defined as 180~\degree - $\theta_{VI}$. $I_{osc,0}$ includes both the currents from the main inverter, and the cross-couple pairs. In a feed-forward ring oscillator, $I_{osc,0}$ also includes the feed-forward current~\cite{Sun2019}. The oscillator current $I_{osc,0}$ is summed with the injection current $I_{inj,0}$, which has its phase defined by $\phi_{0}$. This angle depends on both the injection strength $k_{inj} = |I_{inj}/I_{osc}|$, and the difference between the injection signal frequency \finj and the free-running frequency of the oscillator \ffr. The resulting total current, $I_{t,0}$, has an angle $\psi$ relative to $I_{osc,0}$ and generates a voltage waveform at $V_{90}$, with a phase shift of $\theta_{IV}$ due to first-order RC filtering within the oscillator. For the other stages, they have the phase-shifted injection signals, so the phasor diagram is just a rotated version in Fig.~\ref{fig:phasor}. The phasor relationship can be calculated based on the extended Adler's equation including the amplitude-to-phase conversion~\cite{Adler1946,Wang2025}, or it can be obtained from simulation. Two-stage differential ring oscillators are the simplest oscillator structure, so our analysis starts from it. In the idea case, $V_{0}$ and $V_{90}$ are in perfect quadrature with precise injection signals and delay stages in the oscillator.
%----FIGURE-----
%---------------

\begin{figure}[t!]
\centering 
\includegraphics[width=0.35\textwidth]{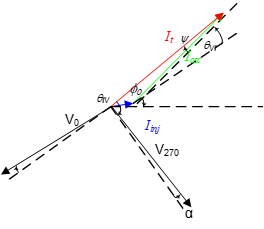}
\vspace{-6pt}
\caption{\small{The phasor diagram of the voltages and currents in an injection-locked ring oscillator with injection phase errors.}}
\label{fig:phasor_dist}
\vspace{-12pt}
\end{figure}

Injection locking ratios are typically designed within the range of 0.05 to 0.2, allowing phase errors in the injection signals to be treated as small-signal disturbances~\cite{Wang2021JSSC,Xu2024}. In a two-stage differential ring oscillator, if there is a phase error $\theta$ at the injection signal $I_{inj,0}$ and $I_{inj,90}$, the common-mode part $\theta/2$ introduce a common-mode phase shift to all the phasors, and the ring oscillator outputs just shift by $\theta/2$. However, the differential phase errors $-\theta/2$ and $\theta/2$ will introduce phase errors at the output. Therefore, phase errors in the injection signal can be analyzed as differential disturbances: $+\theta/2$ at $I_{inj,0}$ and $-\theta/2$ at $I_{inj,90}$. Consequently, voltage signals $V_0$ shifts by $+\alpha$, and voltage signals $V_{90}$ shifts by $-\alpha$, resulting in a quadrature error of $2\alpha$ caused by the input phase error $\theta$. Due to the voltage shift, the current signals $I_{rosc,0}$ and $I_{rosc,90}$ also shift by the amount of $+\alpha$ and $-\alpha$ respectively, and the get summed to the injection signals. This process is illustrated in Fig.~\ref{fig:phasor_dist}.

%----FIGURE-----
%---------------

The height of the new triangle can yield the following equations in the phasor model:

%----EQUATION-----
\begin{equation}
\label{eqa:1}
\begin{split}
& \sin{(\phi_0-\psi-\theta/2+\alpha)}\times|I_{inj}|=\sin{(\psi-2\alpha)}\times|I_{osc}|
\end{split}
\end{equation}
%--------------- 

By differentiating both sides, the sensitivity of the output phase error~($2\alpha$) to the input phase error~($\theta$) can be expressed as:

%----EQUATION-----
\begin{equation}
\label{eqa:2}
\begin{split}
& \frac{d(2\alpha)}{d\theta} = 0.5\frac{|I_{inj}|}{|I_{osc}|}\frac{\cos{(\phi_0+\psi)}}{\cos{\psi}}
\end{split}
\end{equation}
%--------------- 

$\psi$ is the phase shift between total current $I_t$ and the oscillator current $I_{osc}$. Therefore, when the injection ratio $k_{inj}$ is small, $\psi$ should be a small value. From the phasor analysis, we can learn that $psi$ is no greater than $|I_{inj}|/|I_osc|$. For small injection ratios below 0.2, $\psi$ is less than 0.2. Moreover, in a properly biased injection-locked ring oscillator, its free-running frequency is very close to the injection frequency, where $\psi$ approaches zero, allowing sensitivity to be approximated as:

%----EQUATION-----
\begin{equation}
\label{eqa:3}
\begin{split}
& \frac{d(2\alpha)}{d\theta} = 0.5k_{inj}\cos{\phi_0}
\end{split}
\end{equation}
%--------------- 

\begin{figure*}[h]
\centering 
\includegraphics[width=0.8\textwidth]{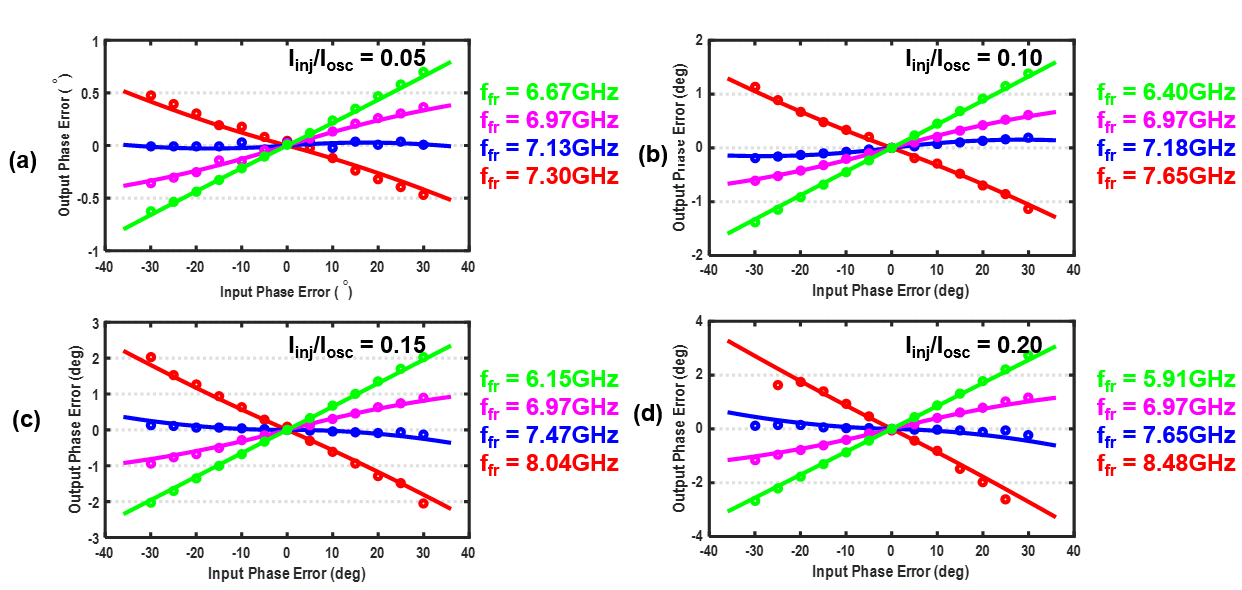}
\vspace{-6pt}
\caption{\small{Calculated and simulated error versus $\phi_0$ input errors under free-running frequencies and injection strength: (a) 0.05; (b) 0.10; (c) 0.15; (d) 0.20.}}
\label{fig:err}
\vspace{-12pt}
\end{figure*}

Fig.~\ref{fig:err} shows the calculated and simulated output phase error versus input phase error for different free-running frequencies and various injection ratios ($k_{inj}=0.05,0.10,0.15,0.20$). The results indicate good agreement. Notably, there exists an optimal frequency where the output phase error becomes completely insensitive to input phase error. This frequency lies slightly above the injection frequency, with values such as 7.13~GHz for $k_{inj}=0.05$, 7.18~GHz for $k_{inj}=0.10$, 7.47~GHz for $k_{inj}=0.15$ and 7.65~GHz for $k_{inj}=0.20$. The optimal jitter point occurs at the center of the locking range~\cite{Adler1946,Razavi2004,Wang2021JSSC}. Consequently, when the input phase error is significant, it is preferable to bias the ring oscillator slightly above the injection frequency to improve performance. Conversely, when the input phase error is minimal, the RO should be biased as close as possible to the injection frequency.

For example, in a cascaded injection-locked oscillator design~\cite{Kinget2002}, this principle is especially relevant. The first stage, which typically provides the most significant phase error correction, should be biased at a slightly higher frequency. Subsequent stages, which further refine the signal, should be biased closer to the injection frequency to achieve optimal jitter performance.

%----FIGURE-----
%---------------

\begin{figure}[t!]
\centering 
\includegraphics[width=0.48\textwidth]{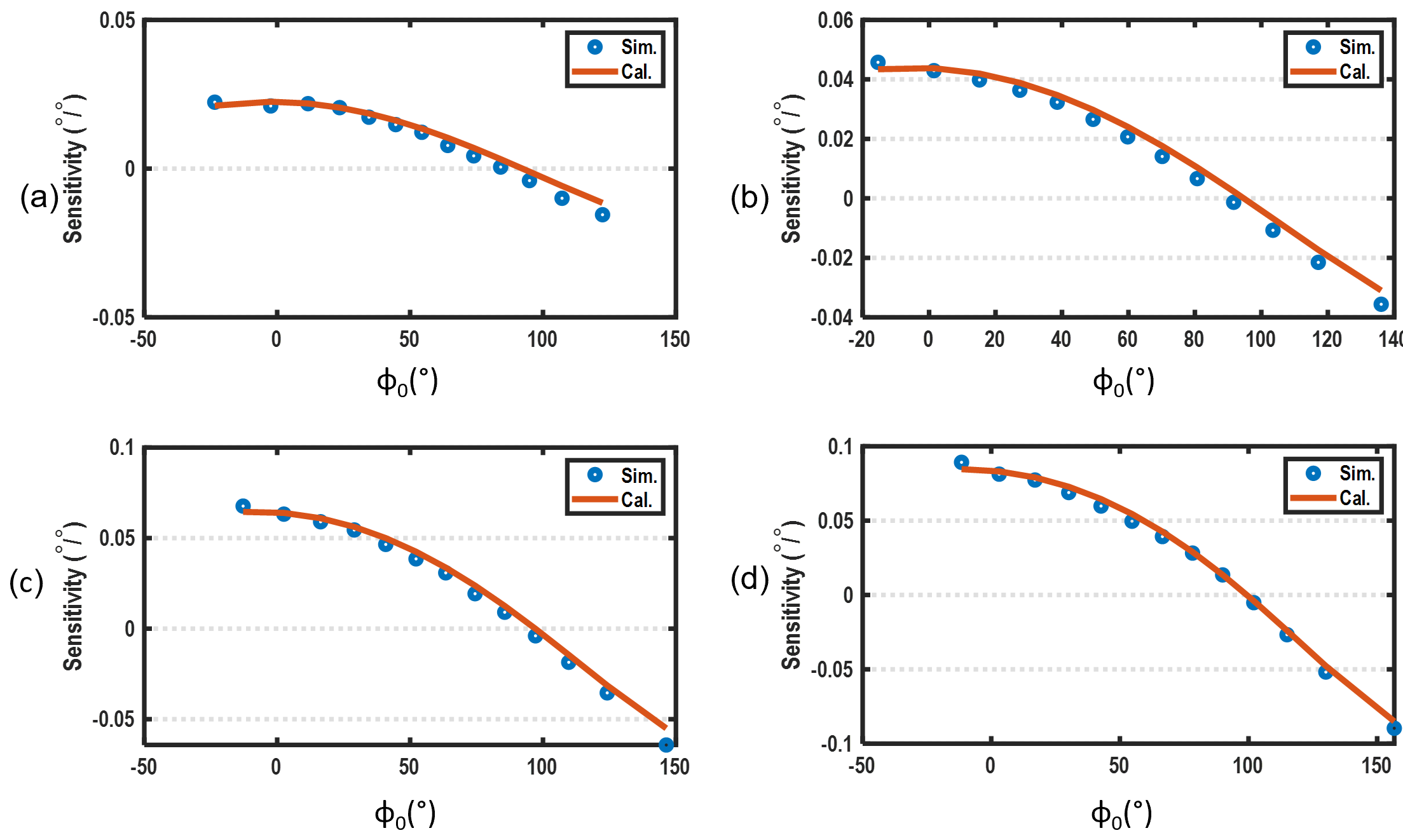}
\vspace{-6pt}
\caption{\small{Calculated and simulated error sensitivity versus $\phi_0$ under injection strength: (a) 0.05; (b) 0.10; (c) 0.15; (d) 0.20.}}
\label{fig:sense}
\vspace{-12pt}
\end{figure}

\begin{figure}[t!]
\centering 
\includegraphics[width=0.48\textwidth]{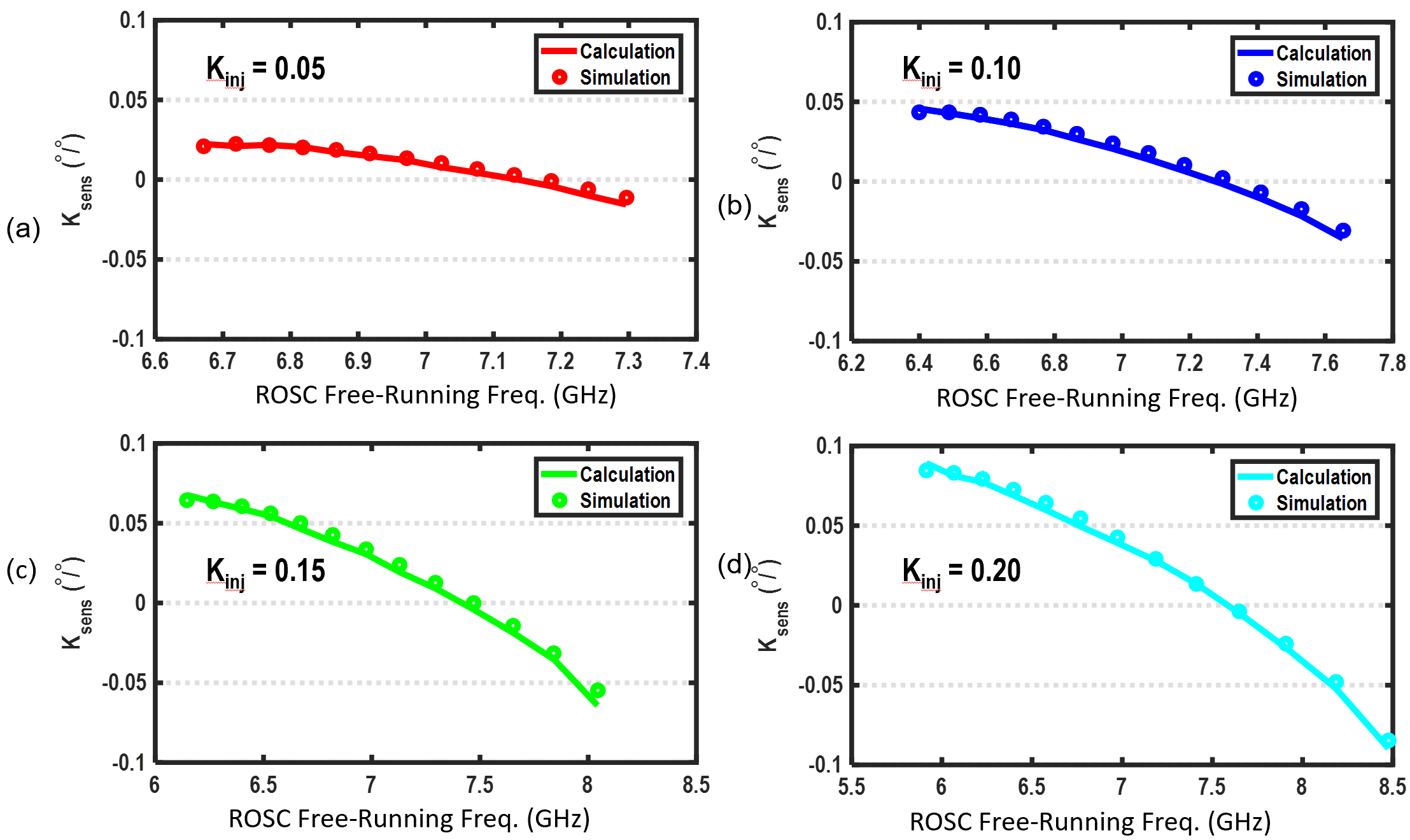}
\vspace{-6pt}
\caption{\small{Calculated and simulated error sensitivity versus free-running frequency under injection strength: (a) 0.05; (b) 0.10; (c) 0.15; (d) 0.20.}}
\label{fig:sense_vs_f}
\vspace{-12pt}
\end{figure}

Fig.~\ref{fig:sense} and Fig.~\ref{fig:sense_vs_f} illustrate the calculated phase error sensitivity as a function of the angle $\phi_0$ and the free-running frequency \ffr, respectively. The results demonstrate that phase error sensitivity depends on both the injection locking ratio and the frequency difference between the injection signal and the free-running frequency. These dependencies align well with the predictions from the analytical model described earlier. When $\phi_0$ reaches 90~\degree, Equation~\ref{eqa:2} reaches close to zero, so the sensitivity approaches zero. Due to the amplitude-to-phase conversion, $\phi_0$ is also affected by the injection amplitude and the summed current amplitude, so it is also a function of the injection strength~\cite{Wang2025}.

%----FIGURE-----
%---------------
%----FIGURE-----
%---------------

The theory can be extended to multi-phase systems by incorporating additional phases into the phasor diagram. For example, in an N-stage differential injection-locked ring oscillator, there are N groups of phasors. However, when there is a phase error in one of the input phases, the differential part should be divided by the number of stages, i.e. $N$. Therefore, for an N-stage ring oscillator, Equation~\ref{eqa:1} and Equation~\ref{eqa:2} should be generalized as:

%----EQUATION-----
\begin{equation}
\label{eqa:1b}
\begin{split}
& \sin{(\phi_0-\psi-\theta/N+\alpha)}\times|I_{inj}|=\sin{(\psi-2\alpha)}\times|I_{osc}|
\end{split}
\end{equation}
%--------------- 

%----EQUATION-----
\begin{equation}
\label{eqa:2b}
\begin{split}
& \frac{d(2\alpha)}{d\theta} = \frac{1}{N}\frac{|I_{inj}|}{|I_{osc}|}\frac{\cos{(\phi_0+\psi)}}{\cos{\psi}}
\end{split}
\end{equation}
%--------------- 

\begin{figure}[t!]
\centering 
\includegraphics[width=0.48\textwidth]{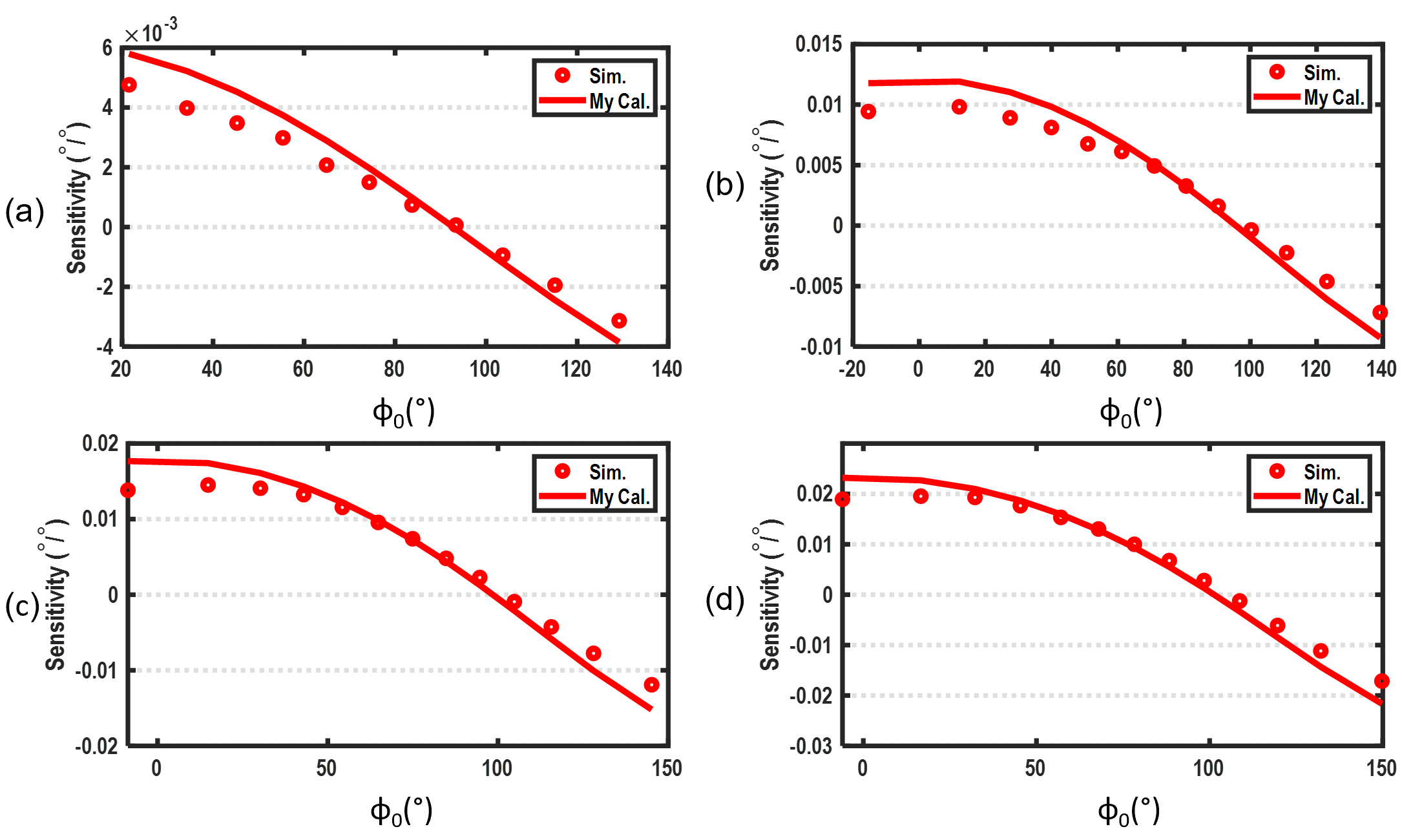}
\vspace{-6pt}
\caption{\small{Calculated and simulated error sensitivity versus $\phi_0$ under injection strength: (a) 0.05; (b) 0.10; (c) 0.15; (d) 0.20.}}
\label{fig:sense_8p}
\vspace{-12pt}
\end{figure}

\begin{figure}[t!]
\centering 
\includegraphics[width=0.48\textwidth]{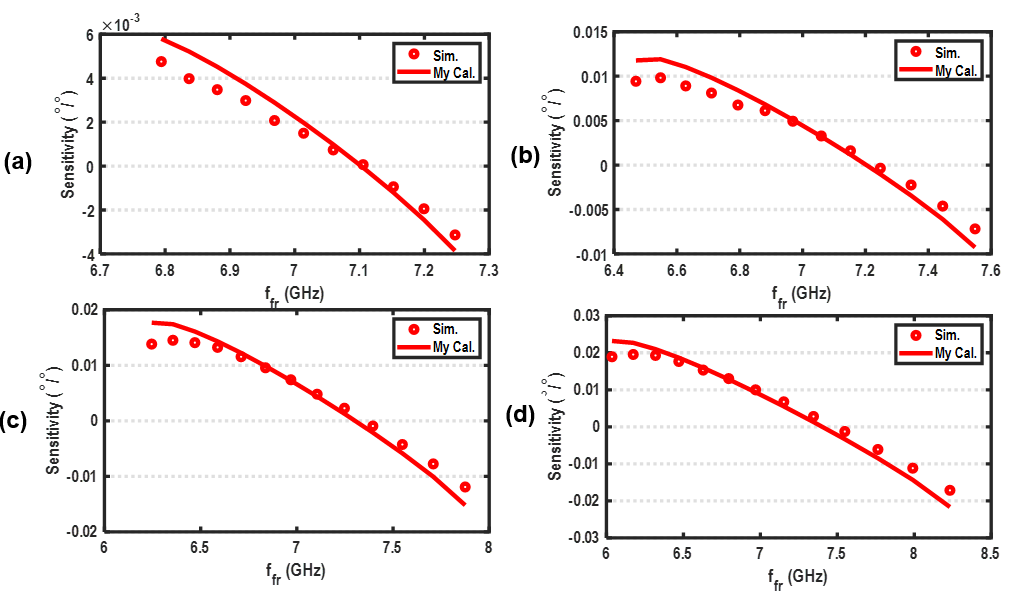}
\vspace{-6pt}
\caption{\small{Calculated and simulated error sensitivity versus free-running frequency under injection strength: (a) 0.05; (b) 0.10; (c) 0.15; (d) 0.20.}}
\label{fig:sense_vs_f_8p}
\vspace{-12pt}
\end{figure}

This also shows that a higher number of stages has better capability of phase correction. Fig.~\ref{fig:sense_8p} and Fig.~\ref{fig:sense_vs_f_8p} present the calculated phase error sensitivity as a function of the angle $\phi_0$ and the free-running frequency \ffr, respectively, for an eight-phase injection-locked ring oscillator~($N=4$). The results confirm that phase error sensitivity is influenced by the injection locking ratio, the frequency offset between the injection signal and the free-running frequency, and the number of stages in the ring oscillator, aligning with predictions from the derived equation. However, this equation is limited to predicting the output errors at the node where the injection signals have errors. As more additional phases are introduced, each stage may experience varying amplitudes and phases, leading to distinct phase errors relative to $\alpha$. Despite this complexity, calculating all precise phase relationships is less practical and meaningful compared to simulation-based analysis. Nevertheless, the equation offers valuable design guidance for biasing the ring oscillator to balance the tradeoff between jitter reduction and phase accuracy improvement.

 %----FIGURE-----
%---------------
%----FIGURE-----
%---------------

\begin{figure}[t!]
\centering 
\includegraphics[width=0.48\textwidth]{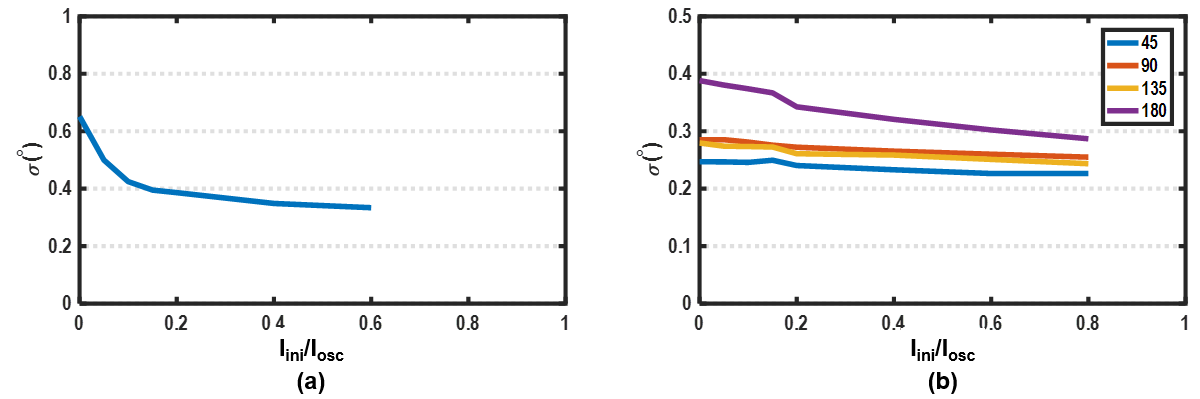}
\vspace{-6pt}
\caption{\small{Simulated random mismatch versus injection strength: (a) a two-stage differential RO; (b) a four-stage differential RO.}}
\label{fig:monte}
\vspace{-12pt}
\end{figure}

Moreover, injection locking effectively mitigates random errors in the injection-locked oscillator. Therefore, when the ILO is highly sensitive to its own random mismatches as technology nodes scale aggressively and oscillators operate at higher frequencies, it is recommended to use a stronger injection locking ratio to suppress these errors. Fig.~\ref{fig:monte} illustrates the relationship between random mismatch and the injection ratio. As demonstrated, increasing the injection ratio help reduce the random mismatch.

%----FIGURE-----
%---------------
%======================================================================
\section{Conclusions}
\label{sec:conclusions}
%======================================================================
This paper introduces a phasor model for analyzing phase accuracy in multi-phase injection-locked ring oscillators. Simulation results demonstrate excellent alignment with the analytical predictions, validating the model's accuracy. The study reveals that the optimal frequency for phase improvement is slightly higher than the injection frequency. Additionally, injection locking reduces random mismatches in the injection-locked oscillator. Consequently, there exists an optimal injection strength in practical applications, considering factors such as jitter, input errors from the injection signals, and phase errors inherent to the ILO.

%====================================================================== \section*{Acknowledgments}

%======================================================================

%======================================================================
% References
%======================================================================
%\clearpage
\bibliographystyle{IEEEtran}
%\bibliography{IEEEabrv,Journal_ILO}
\bibliography{Journal_ILO}
\end{document}